\documentclass{aa}  

\usepackage{graphicx}
\usepackage{amsmath}	% Advanced maths commands
\usepackage{natbib}
\usepackage{subcaption}
\usepackage{booktabs}    % horizontal line in multicolumn tab
\usepackage{comment}
\usepackage{txfonts}
\usepackage{xspace}
\newcommand{\s}{CXOU J110926.4--650224\xspace}
\newcommand{\fermis}{4FGL\,J1110.3--6501\xspace}
\newcommand{\unitL}{erg~s$^{-1}$\xspace}

\newcommand{\fermi}{\textit{Fermi}\xspace}
\newcommand{\chandra}{\textit{Chandra}\xspace}
\def\deg{\mbox{$^{\circ}$}}

\usepackage{natbib,twoopt}
\usepackage[breaklinks, colorlinks]{hyperref} 
\usepackage{xcolor}
\usepackage[hyphenbreaks]{breakurl}
\DeclareMathAlphabet\mathzapf       {T1}{pzc} {mb} {it}
\bibpunct{(}{)}{;}{a}{}{,}             
\definecolor{cobalt}{rgb}{0.06, 0.2, 0.65}
\hypersetup{
  colorlinks,
  citecolor=cobalt,
  linkcolor=cobalt,
  urlcolor=cobalt,
}
\makeatletter
  \newcommandtwoopt{\citeads}[3][][]{\href{http://adsabs.harvard.edu/abs/#3}%
    {\def\hyper@linkstart##1##2{}%
     \let\hyper@linkend\@empty\citealp[#1][#2]{#3}}}
  \newcommandtwoopt{\citepads}[3][][]{\href{http://adsabs.harvard.edu/abs/#3}%
    {\def\hyper@linkstart##1##2{}%
     \let\hyper@linkend\@empty\citep[#1][#2]{#3}}}
  \newcommandtwoopt{\citetads}[3][][]{\href{http://adsabs.harvard.edu/abs/#3}%
    {\def\hyper@linkstart##1##2{}%
     \let\hyper@linkend\@empty\citet[#1][#2]{#3}}}
  \newcommandtwoopt{\citeyearads}[3][][]%
    {\href{http://adsabs.harvard.edu/abs/#3}
    {\def\hyper@linkstart##1##2{}%
     \let\hyper@linkend\@empty\citeyear[#1][#2]{#3}}}
\makeatother
\usepackage[switch]{lineno}
\begin{document}

   \title{Identification and characterisation of the gamma-ray counterpart of the transitional pulsar candidate \s}
   \titlerunning{Fifteen years of \fermi-LAT observations of \s}

    \author{A.~Manca
          \inst{1}\fnmsep\thanks{E-mail: arianna.manca@inaf.it}
          \and 
             F.~Coti Zelati\inst{2,3}
          \and
            J.~Li\inst{4,5}
          \and
            D.~F.~Torres\inst{2,3,6}
          \and
            J.~Ballet\inst{7}
          \and
            A.~Marino\inst{2,3}
          \and
            A.~Sanna\inst{1}
          \and
            N.~Rea\inst{2,3}
          \and
            T.~Di~Salvo\inst{8,9,10}
          \and
            A.~Riggio\inst{1,10}
          \and
            L.~Burderi\inst{1,11}
          \and
            R.~Iaria\inst{8}
          }

   \institute{Universit\`a degli Studi di Cagliari, Dipartimento di Fisica, SP Monserrato-Sestu km 0.7, I-09042 Monserrato, Italy
         \and
            Institute of Space Sciences (ICE, CSIC), Campus UAB, Carrer de Can Magrans s/n, E-08193 Barcelona, Spain
        \and
            Institut d'Estudis Espacials de Catalunya (IEEC), 08860 Castelldefels (Barcelona), Spain
        \and
            CAS Key Laboratory for Research in Galaxies and Cosmology, Department of Astronomy, University of Science and Technology of China, Hefei, PR China
        \and
             School of Astronomy and Space Science, University of Science and Technology of China, Hefei, PR China
        \and
            Institució Catalana de Recerca i Estudis Avançats (ICREA), E-08010 Barcelona, Spain
        \and
            Université Paris-Saclay, Université Paris Cité, CEA, CNRS, AIM, 91191, Gif-sur-Yvette, France
        \and
             Universit\`a degli Studi di Palermo, Dipartimento di Fisica e Chimica - Emilio Segrè, via Archirafi 36 - 90123 Palermo, Italy
         \and
            INFN, Sezione di Cagliari, Cittadella Universitaria, 09042 Monserrato, CA, Italy
        \and
            INAF - Osservatorio Astronomico di Cagliari, via della Scienza 5, 09047 Selargius (CA), Italy
        \and
            INAF/IASF Palermo, via Ugo La Malfa 153, I-90146 - Palermo, Italy
             }

   \date{Received XXX; accepted XXX}

% \abstract{}{}{}{}{} 
% 5 {} token are mandatory
 
  \abstract
   {Transitional millisecond pulsars (tMSPs) represent a crucial link between the rotation-powered and accretion-powered states of binary pulsars. During their active X-ray state, tMSPs are the only low-mass X-ray binary systems detected up to GeV energies by the \fermi Large Area Telescope (LAT). \s\ is a newly discovered tMSP candidate in an active X-ray state, potentially spatially compatible with a faint gamma-ray source listed in the latest \fermi-LAT point-source catalogue as \fermis. Confirming the association between \s\ and the \fermi\ source is a key step toward validating its classification as a tMSP. In this study, we analyse \fermi-LAT data collected from August 2008 to June 2023 to achieve a more accurate localisation of the gamma-ray source, characterise its spectral properties, and investigate potential time variability. By thoroughly reconstructing the gamma-ray background around the source using a weighted likelihood model, we obtain a new localisation that aligns with the position of the X-ray source at the 95\% confidence level, with a Test Statistic value of $\sim 42$. This establishes a spatial association between the gamma-ray source and \s. The gamma-ray emission is adequately described by a power-law model with a photon index of $\Gamma = 2.5 \pm 0.1$ and a corresponding flux of $(3.7\pm0.9) \times 10^{-12}$\,erg\,cm$^{-2}$\,s$^{-1}$ in the 0.1--300\,GeV range.}


   \maketitle
%
%-------------------------------------------------------------------

\section{Introduction}
Millisecond pulsars (MSPs) are a subcategory of pulsars characterised by very short ($\lesssim$~10~ms) rotation periods and low ($\sim$10$^{8-9}$~G) magnetic fields. %The low magnetic fields are typical of old pulsars that have undergone magnetic field decay. 
The existence of MSPs provides evidence for the so-called recycling scenario of pulsars \citep{Alpar82}: these objects are thought to be the outcome of a phase of accretion within a Low-Mass X-ray Binary (LMXB) system, which causes the neutron star (NS) to spin up to millisecond periods (accretion-powered state). When the accretion phase ends, after $\sim 10^8-10^9$~years, the NS becomes detectable as a radio pulsar powered by the rotation of its magnetic field (rotation-powered state).

Transitional millisecond pulsars (tMSPs) are sources switching between the accretion- and rotation-powered states, representing the link between accreting X-ray pulsars and evolved radio MSPs, and thus providing strong evidence for the recycling scenario. Only three tMSPs have been identified up to now and only one has been observed in a full accretion outburst state (for a review, see \citealt{PapittoDeMartino2022}).
%each showing slightly different observational properties. 
The tMSPs exhibit three distinct states depending on the mass transfer rate from the companion -- a key parameter that regulates the equilibrium between the gravitational attraction on the matter and the outward pressure exerted by the pulsar wind. The first is an accretion outburst state with X-ray luminosity of $\sim 10^{36-37}$~\unitL \citep[only observed in IGR J18245--2452,][]{Papitto2013swings, Ferrigno2014}; the second is a radio pulsar state (L$_X \lesssim 10^{33}$~\unitL); the third is an intermediate state characterised by the presence of an accretion disc and an X-ray luminosity of $10^{33-34}$~\unitL, i.e. in between that measured in the accretion state and the radio pulsar state. This state is referred to as "sub-luminous disc state" or "active X-ray state" and was observed in both PSR J1023+0038 and XSS J12270--4859 \citep[see, e.g.,][]{Linares2014}. Three distinct intensity modes can be identified in this intermediate state, as observed in the X-ray light curves: the high mode, characterised by a power-law spectrum with photon index of $\Gamma \sim 1.7$ and luminosity of $L_X \approx 10^{33}$~\unitL; %, and no modulation at the orbital period; 
the low mode, characterised by a slightly softer power-law spectrum with $\Gamma \sim 1.8$ and luminosity of $L_X \approx 10^{32}$~\unitL; and the flaring mode, characterised by sudden increases in count rates on timescales of less than one minute to hours, which can reach luminosities of $\approx$10$^{34}$~\unitL \citep{Tendulkar2014}. The switches between modes can occur in about 10\,s.

tMSPs have been observed across several bands, ranging from gamma-ray to radio frequencies. Among these, gamma-ray emission offers valuable insights into the processes occurring in tMSPs and is one of their distinctive features. %This emission enables the identification of additional candidate sources by examining their gamma-ray properties. 
Emission at such high energies implies active non-thermal processes at play. In the sub-luminous state of confirmed tMSPs, the gamma-ray emission is several times more intense than that observed in the rotation-powered state, indicating an enhancement of these non-thermal processes. For instance, PSR J1023+0038 exhibits gamma-ray emission that is around five times brighter in its sub-luminous state compared to its rotation-powered state \citep{Stappers2014}. Notably, similar levels of gamma-ray emission have not been detected in other LMXBs (neither in quiescence nor during accretion phases), making tMSPs the only LMXBs confirmed to emit in the GeV range (though some LMXB systems have potential gamma-ray counterparts, e.g., \citealt{OnaWilhelmi2016}, requiring further evidence for definitive association). The gamma-ray emission of tMSPs in the sub-luminous state is characterised by a power-law spectrum with a photon index of $\Gamma \sim 2$ and a cut-off energy between 4 and 10\,GeV, as observed in both PSR J1023+0038 and XSS J12270--4859 \citep{Torres2017, Torres2022}.

The enhanced gamma-ray emission of tMSPs in their sub-luminous state  raises questions about its origin -- whether it results from the combined gamma-ray radiation of a central rotation-powered pulsar and an additional non-thermal mechanism, or if it arises entirely from a novel non-thermal process. %because tMSPs are the only LMXBs detected at energies extending into the GeV range. 
Two primary scenarios have been proposed to explain the origin of this gamma-ray emission. In the propeller scenario, the ejected matter emits synchrotron radiation, which is subsequently up-scattered to gamma-ray energies through inverse Compton scattering (ICS) in the boundary region where the accretion disc is truncated by the neutron star magnetosphere \citep{Papitto2014, Papitto2015}. In the mini-pulsar wind nebula model, the rotation-powered pulsar emits a pulsar wind that interacts with the inner regions of the accretion disc, creating a shock. This shock emits radiation in the X-ray, UV and optical bands. The synchrotron emission produced at the shock is then up-scattered by ICS of UV photons from the disc by the pulsar wind, producing detectable GeV emission. This model accounts for the periodic pulsations observed in the X-ray, UV, and optical emissions, as the shock emission is modulated at the pulsar spin period \citep{Papitto2019, Veledina2019, MiravalZanon2022, Illiano2023}.

Candidate tMSPs are efficiently identified by searching for counterparts of unidentified gamma-ray sources that exhibit time variability in the X-rays, analogous to the high and low modes characteristic of tMSPs, and possess a spectral shape consistent with a power-law model with a photon index of $\Gamma \sim 1.7$. Additionally, tMSPs display peculiar double-peaked emission lines in the optical band, which arise from the presence of an accretion disc \citep[see, e.g.,][for references on 1RXS J154439.4--112820]{BogdanovHalpern2015, Bogdanov2016}.

The candidate tMSP \s was first detected by \chandra in 2008 and identified as the soft X-ray counterpart of the hard X-ray source IGR J11098--6457 \citep{Tomsick2009}. The source showed a behaviour typical of tMSPs in the sub-luminous state, with a bimodal variability in the soft X-ray emission and the presence of a hard X-ray counterpart \citep{CotiZelati2019}. 
%The X-ray count rates showed a bimodal distribution with occasional flares, and the spectrum was well-fitted by a power-law model with $\Gamma = 1.63 \pm 0.01$. 
An optical counterpart, of magnitude $G \sim 20.1$, was also identified, at an estimated distance of $\approx 4$~kpc. 
Subsequent coordinated X-ray and radio observations detected a radio counterpart with an average flux density of $\sim$33\,$\mu$Jy at 1.28\,GHz, which exhibited significant variability and occasional flaring events shortly after X-ray flares. However, no clear correlated or anti-correlated variability was observed between the X-ray and radio emissions \citep{CotiZelati2021}. More recently, high-time-resolution X-ray, optical, and near-infrared observations revealed that the variability patterns of the X-ray and optical emissions are strongly correlated and that the near-infrared emission was strongly variable. Spectroscopic analyses suggest that the companion star is of spectral type between K0 and K5 \citep{CotiZelati2024}. 
%See \citet{CotiZelati2021, CotiZelati2024} for extensive discussion of multi-wavelength searches.

\begin{figure}
	\includegraphics[width=1.0\columnwidth]{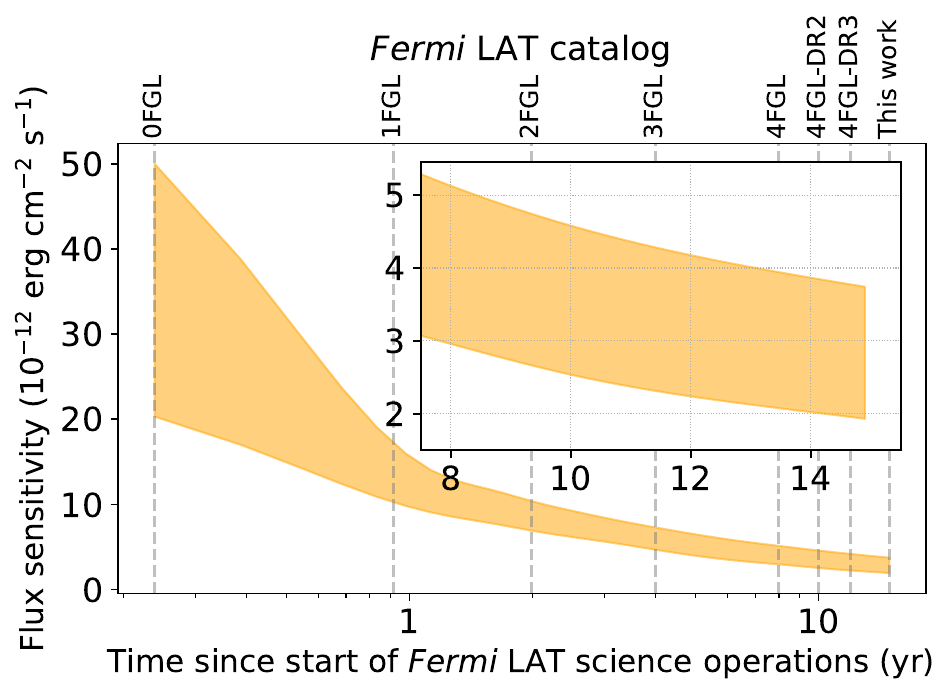}
    \caption{Sensitivity to detectable gamma-ray emission by the \fermi-LAT at the position of \fermis as a function of time. The inset shows a zoomed-in view for the most recent time span. Flux values have been estimated assuming a Test Statistic (TS) threshold of 25, performing a scan over photon indices ranging from 1 to 5 to mimic various spectral scenarios. The calculation accounts for the local background and exposure at the source position.}
    \label{fig:sensitivity}
\end{figure}

A gamma-ray source with a position that is fully compatible with that of \s\ was initially detected by \fermi Large Area Telescope (LAT) based on 8 years of data collection\footnote{\url{https://fermi.gsfc.nasa.gov/ssc/data/access/lat/fl8y/}}. %After a spatially coincident source was identified in the FL8Y catalogue, \s remained undetected in both the 4FGL catalogue \citep{4FGL} and the second data release (4FGL-DR2, \citealt{4FGL-DR2}). The %refined position derived utilising 12 years of data collection (DR4) is formally not compatible with that of \s at the 95\% c.l. (see \citealt{Fermi4FGL} and \citealt{FermiDR4} for more details on the latest release of the \fermi catalogue).
However, it was not detected in the first 4FGL catalogue \citep{Fermi4FGL} or in the second data release (4FGL-DR2; \citealt{4FGL-DR2}). A gamma-ray source, \fermis, was detected again in the 4FGL-DR3 and DR4 data releases (see \citealt{Fermi4FGL-DR3} and \citealt{FermiDR4} for more details on the latest release of the \fermi catalogue). However, the position of the 4FGL-DR3 and DR4, derived from 12 years of data collection, is not compatible with that of \s at the 95\% confidence level (c.l.).

In this paper, we analyse about 15 years of \fermi-LAT data in order to establish a connection between the X-ray and gamma-ray sources. The paper is organised as follows: in Section\,\ref{sec:analysis}, we describe the \fermi-LAT data analysis methods used to investigate the gamma-ray emission from \fermis. Section\,\ref{sec:results} presents the results of our analysis, including source localisation and spectral characterisation. In Section\,\ref{sec:discussion}, we discuss the implications of our findings within the context of tMSPs. Finally, Section\,\ref{sec:conclusions} summarises our conclusions and outlines prospects for future research.
%We conduct a detailed modelling of the gamma-ray background to determine a precise localisation and derive an estimation of its gamma-ray flux.

\begin{figure}
    \centering
    \includegraphics[width=0.45\textwidth]{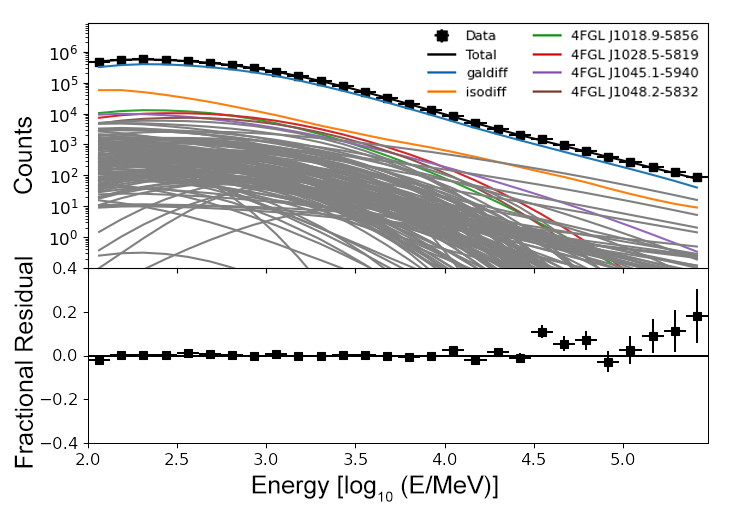}
    \caption{Count spectrum and residuals resulting from the likelihood fit of the data within the considered RoI. The fit was performed using a total of 184 sources (see the text for details). The black line represents the total model, where the predominant component is the Galactic diffuse emission (indicated in azure), followed by the isotropic background gamma-ray emission component (represented in orange). Some of the brightest sources in our RoI are also highlighted.}
    \label{fig:spectrum}
\end{figure}

\section{Data analysis}
\label{sec:analysis}
In our analysis, we use data collected by the \fermi-LAT between August 5, 2008 (MJD 54683) and June 29, 2023 (MJD 60124), approximately 14 years and 11 months in total. Our analysis was restricted to events located within a circular region of interest (RoI) with a radius of 15\deg, centred on the nominal position of the \fermi source \fermis (RA = 11$^\mathrm{h}$10$^\mathrm{m}$16$^\mathrm{s}$, DEC = -65$^{\circ}$01$^{\prime}$06$^{\prime\prime}$; see \citealt{Fermi4FGL}). We considered events with energies ranging from 0.1 to 300\,GeV.

The data extraction and analysis were performed using the \textsc{Fermitools} software package \citep[v2.2.0;][]{Fermitools2019} and the \textsc{Fermipy} python suite \citep[v. 1.3.1;][]{Fermipy2017}. We selected the source events separating between front and back events of the LAT tracker and having zenith angle $<$90\deg\ so as to reduce contamination from the Earth limb. Corresponding good time intervals were created with the \texttt{gtmktime} tool.

We computed the gamma-ray flux and spectral parameters of \fermis via a joint weighted binned maximum likelihood fitting technique \citep{Fermi4FGL}, adopting the latest instrument response function, "P8R3$_-$SOURCE$_-$V3". 
Starting from our filtered data, we created a square counts map of size 21.2\deg, with an image scale of 0.1\deg\,pix$^{-1}$ and AIT \citep[Aitoff,][]{Calabretta2002} projection method, allowing us to visualise the Galactic diffuse emission component.
We estimated the LAT flux sensitivity at the position of \fermis to be $\simeq(2-4)\times10^{-12}$~erg~cm$^{-2}$~s$^{-1}$ (see Figure\,\ref{fig:sensitivity}).

%We also created a counts cube, in order to observe how the counts map varies with energy. We created 35 logarithmically spaced maps in the energy range 100 MeV -- 300 GeV.
%We derived a livetime cube (\texttt{gtltcube}) to take into account the different time intervals during which the source was observed at specific inclination angles with respect to the normal of the instrument. The cube was created mapping the ROI with step size cos($\theta$) = 0.025 and 1 degree pixel size.

One key aspect of the analysis involves accurately modelling the background gamma-ray emission in a sky region that is densely populated with sources, as is the case in our study. We used the \texttt{make4FGLxml} script\footnote{\url{https://github.com/physicsranger/make4FGLxml}} to generate a spatial-spectral model that includes the model for the Galactic interstellar diffuse gamma-ray emission ("\textit{gll\_iem\_v07.fits}"), the spectral template for the extragalactic isotropic background emission  ("\textit{iso\_P8R3\_SOURCE\_V3\_v1.txt}"), and all the sources listed in the latest \fermi-LAT data release (4FGL-DR4) located within the RoI.
%utilizing the LAT catalogue file \textit{gll\_psc\_v28.xml}. 
Sources listed in the 4FGL-DR4 and lying within a square region of width 25\deg\ centred on the target source position were also added in the modelling. The emission from these sources is described using the best-fitting models derived in the 4FGL-DR4, such as a power-law, a LogParabola (usually adopted for blazars), and a power-law with a high-energy exponential cutoff (typical for pulsars). The model includes 184 sources (174 point-like sources and 10 extended sources).
%Before starting the fit, we made sure to run the \texttt{optimize()} tool, to ensure the fit starting points were as accurate as possible.
To enhance accuracy of our starting parameters for the fit, we executed the \texttt{optimize()} tool prior to initiating the fitting process.

%We created an exposure cube (\texttt{gtexpcube2}) where each map is 212x212 pixels, and image scale 0.1. We directly passed the instrument response function (IRF) and kept all the selections as we did for the previous analysis steps, considering 35 energy bins. We also created an all-sky version of the exposure cube, with image size 1800x900 pixels and image scale 0.1.

%We computed the model counts map with \texttt{gtsrcmaps}, considering previously produced files and the CALDB calibration files. In order to include all the background sources in the region of interest, we used the all-sky exposure cube for the computation.

We performed a joint weighted likelihood fit employing the \texttt{gtlike} tool with the \textsc{newminuit} optimisation algorithm. % utilizing the all-sky source map and exposure cubes and 
We derived the weighted maps from our filtered files and used them in the likelihood fit\footnote{\url{https://fermi.gsfc.nasa.gov/ssc/data/analysis/scitools/weighted_like.pdf}}.
In the fitting process, we allowed the normalisations of all sources within 2\deg\ of the target position to vary, along with the parameters of the Galactic and isotropic background components, as well as the parameters of our target (which is modelled using a power-law). 
%The tool creates an output model based on the best-fitting parameters. 
%After completing the analysis, we still observed a residual tail in the fit at energies $\gtrsim$10\,GeV, as shown in the residuals of the model compared to the counts spectrum (Figure\,\ref{fig:spectrum}). %[MOVED]We assessed the goodness-of-fit using a PS map, as described by \cite{Bruel2021}. From the weighted PS map, we derived the histogram of the PS values (Figure\,\ref{fig:histo-PS}), which indicates the presence of some unmodelled residuals. However, considering the densely populated nature of the region under study and the steady level of counts in the residual excess map near our source (Figure\,\ref{fig:excess}), the model provides an acceptable representation of the background emission.

\begin{figure*}
    \centering
    \begin{subfigure}{0.46\textwidth}
        \includegraphics[width=\textwidth]{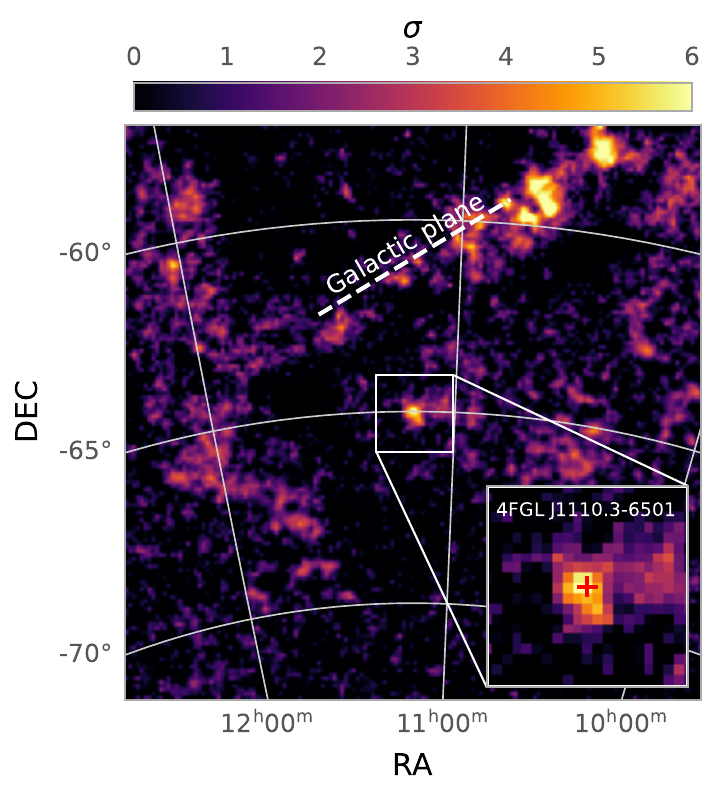}
        \caption{TS map}
        \label{fig:TS}
    \end{subfigure}
    %\hspace{-1.2cm}
    \begin{subfigure}{0.48\textwidth}
        \includegraphics[width=\textwidth]{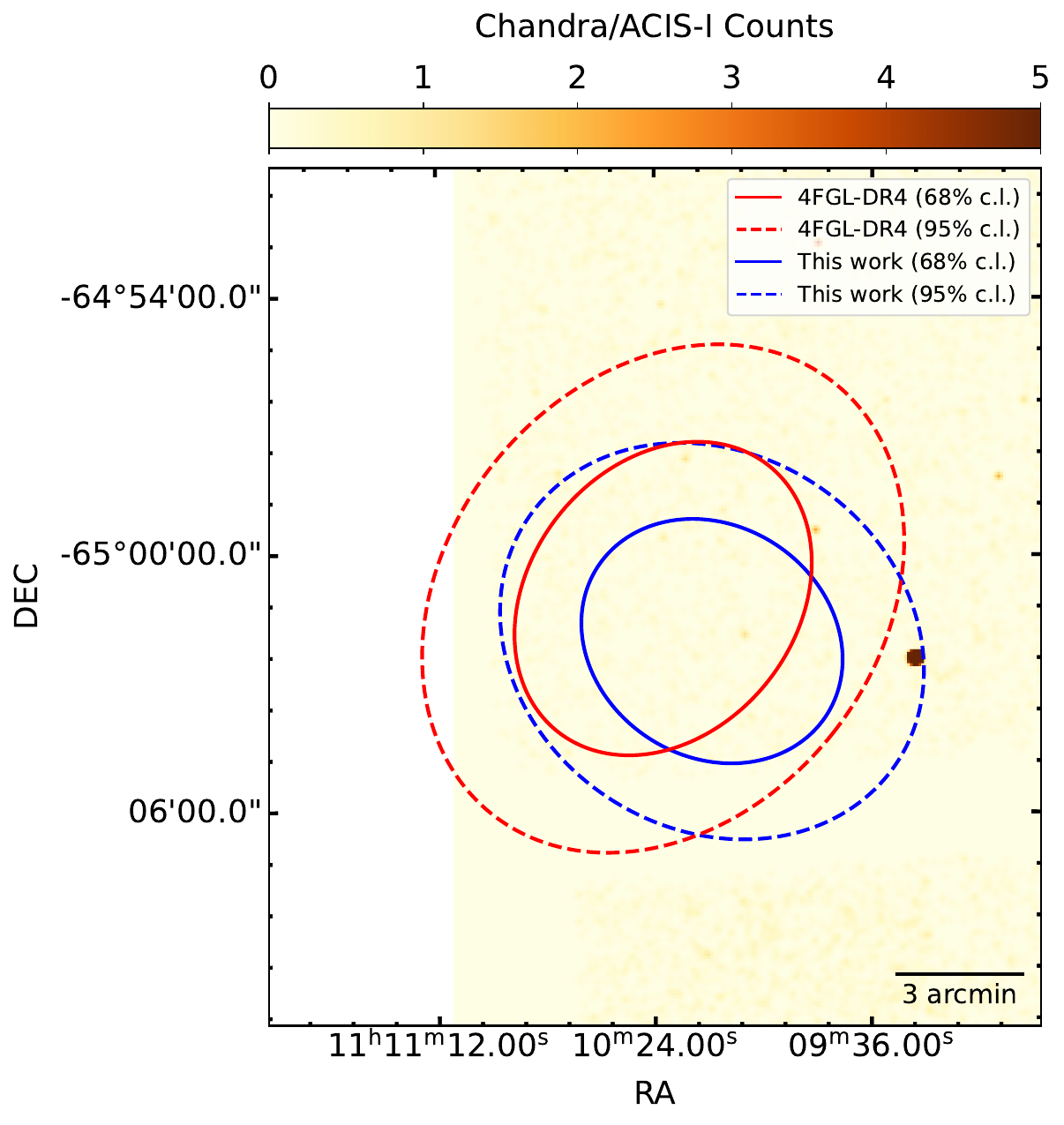}
        \caption{Localisation}
        \label{fig:loc}
    \end{subfigure}
    
    \caption{\textit{Left}: TS map of the RoI. Regions with significant, unmodelled contributions are shown in brighter colours. The dashed line indicates the position of the Galactic plane. In the centre of the RoI we can identify our source as the yellow spot, as depicted in the inset. 
    \textit{Right}: Comparison of our source localisation with the previous \fermi 4FGL-DR4 catalogue position, overlaid onto the \chandra X-ray image. The catalogue position is represented by red ellipses, while the position from our study is shown in blue. The solid ellipses indicate the 68\% c.l., while the dashed ellipses indicate the 95\% c.l. The \chandra source \s, depicted as the dark spot, lies within the 95\% c.l. contour, consistent with our localisation of the \fermi source.}
    \label{fig:TS_loc}
    
\end{figure*}

%We assessed the goodness-of-fit using a PS map, as described by \citep{Bruel2021}. 
To determine the significance of the source, we used the Test Statistic (TS), defined as \\citep{Mattox1996}:

\begin{equation}
    {\rm TS} = 2 (\log \mathcal{L}_1 - \log \mathcal{L}_0),
    \label{eq:TS}
\end{equation}
where log$\mathcal{L}_1$ represents the likelihood including our source in the model, and log$\mathcal{L}_0$ represents the likelihood without our source. %Consequently, we executed the fitting procedure twice to compare these scenarios: once with our source included in the model, and once without it.

%The fitting tool does not vary the localisation of the source. 
%To hone the source position, we followed a two-step process. 
%Firstly, we generated a TS map on a finite grid and determine the initial localisation using the highest TS value on the map. Next, we honed the localisation by maximising the TS value using a more precise grid of values. 
%Firstly, we generated a likelihood map of the region to determine the initial localisation. Next, we performed a scan of the likelihood surface within a search region centered on the initial estimate and encompassing the 99\% positional uncertainty contour. This operation is accomplished with the \texttt{localize()} method.
We generated a TS map using the \texttt{gttsmap} tool (Figure \ref{fig:TS}), excluding our source to highlight it as an excess. Next, we determined the gamma-ray source position with the \texttt{localize()} tool, which employs a TS map to determine the source position through a likelihood fit. Finally, we ran \texttt{gtlike} again at the newly determined position to obtain the final spectral results.% (Figure \ref{fig:loc}).}

%We then extracted a light curve by splitting the data into four equal intervals, each lasting 1360 days. We applied the previously mentioned fitting procedure to each of these time intervals using the \texttt{lightcurve()} method. The decision to use four bins for extracting the light curve aims to increase the number of detections (particularly given the faint gamma-ray emission from the source) as well as to reduce the uncertainties on the single measurements and investigate the presence of any potential time variability. 
%This allowed us to determine the flux at each time bin. 
%We then extracted a Spectral Energy Distribution (SED) using three equally spaced logarithmic energy intervals (\texttt{sed()} method), restricting the energy range to 1--10~GeV, where the source is detected at higher significance. 
\begin{figure*}
    \centering
    \begin{subfigure}{0.4\textwidth}
        \includegraphics[width=\textwidth]{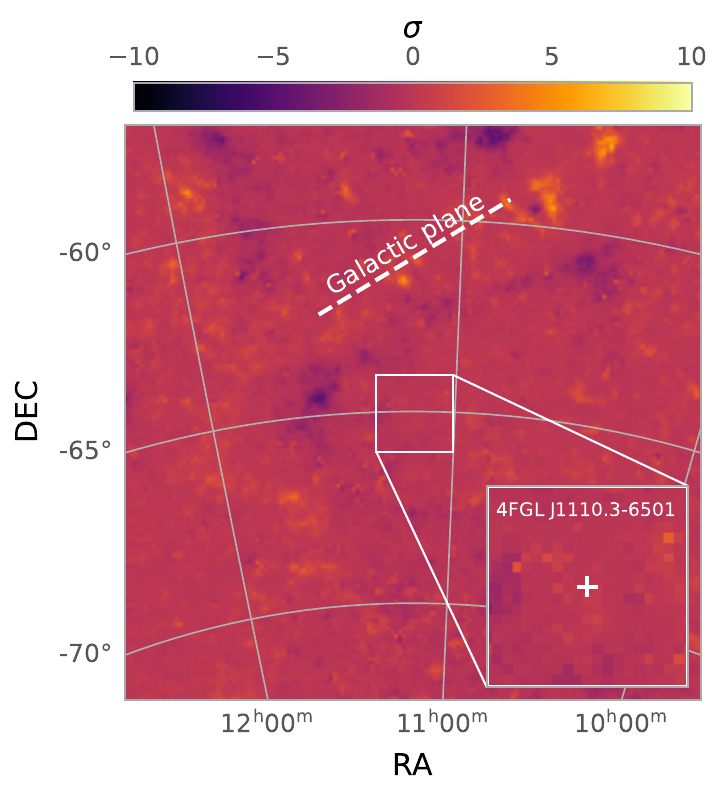}
        \caption{PS map of the RoI.}
        \label{fig:PSmap}
    \end{subfigure}
    \begin{subfigure}{0.55\textwidth}
        \includegraphics[width=\textwidth]{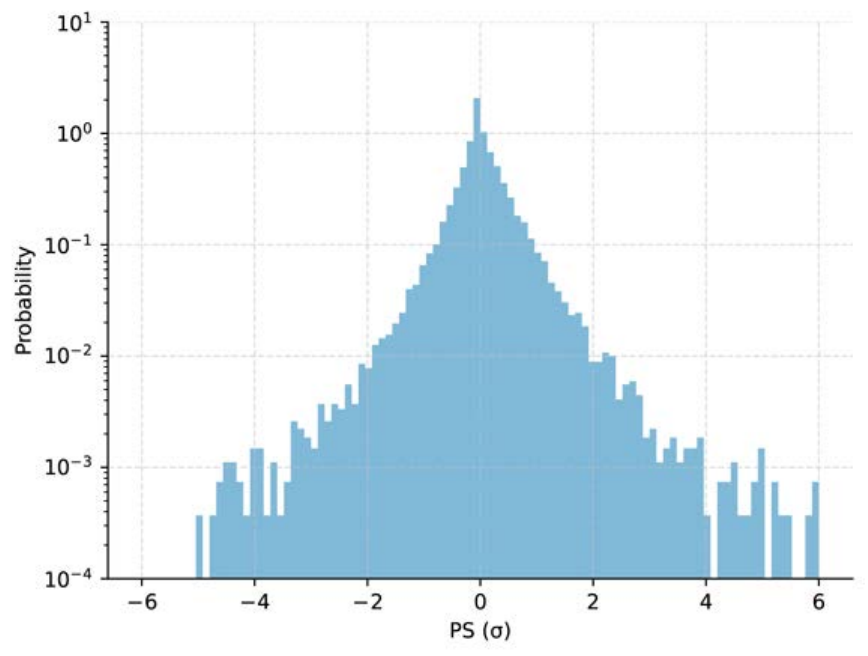}
        \caption{Histogram of the PS in units of $\sigma$}
        \label{fig:histo-PS}
    \end{subfigure} 
    \caption{\textit{Left panel}: PS map of the RoI. The gradient represents positive and negative residuals in the RoI, rescaled in terms of statistical significance according to the definition in \citet{Bruel2021}. Negative values imply an overestimation of the background. The Galactic plane is highlighted with a dashed line. In the inset, the white cross shows the position of our source, included in the model and therefore blended with the background. \textit{Right panel}: Histogram of the PS values in units of standard deviations, derived from the joint weighted likelihood fit that includes the source under examination in the spectral model. Negative values correspond to areas where the model overestimates the background, while positive values indicate areas where the model underestimates it.}
    \label{fig:PS}
\end{figure*}
\section{Results}
\label{sec:results}
Figure\,\ref{fig:spectrum} presents the count spectrum and residuals resulting from the joint likelihood fit of the background model and the target source emission, as a function of energy. At higher energies, starting from $\approx$10\,GeV, we observe residuals that were not accurately accounted for by the fitting process. Nevertheless, these residuals exhibit relatively small scatter and occur in an energy range where significant emission from \fermis is not expected. The dominant component in the spectrum is the Galactic diffuse emission (shown by the azure line), followed by the isotropic background emission component (depicted by the orange line). We have highlighted the contributions from four of the brightest sources in our RoI in the \fermi\ 4FGL-DR4 catalogue: 4FGL J1018.9--5856, 4FGL J1028.5--5819, 4FGL J1045.1--5940, and 4FGL J1048.2--5832, all located along the Galactic plane. The remaining sources in our model are represented by grey lines.

Figure \ref{fig:TS} presents the TS map of the RoI. Bright residuals are evident along the Galactic plane. Our source is detected with a TS value of approximately 42 (calculated as defined in Equation \ref{eq:TS}).%, corresponding to a significance of 4.7 $\sigma$. 
The best-fitting photon index is $\Gamma = 2.5 \pm 0.1$, and the flux is $(3.7 \pm 0.9) \times 10^{-12}$\,erg\,cm$^{-2}$\,s$^{-1}$ in the 0.1--300\,GeV energy range.

Figure\,\ref{fig:loc} compares the localisation provided in the \fermi\ 4FGL-DR4 catalogue (red ellipses) with the updated localisation derived from our analysis (blue ellipses; see Table\,\ref{tab:loc} for details). These are superimposed on an archival \chandra X-ray image of the field. Although \s\ falls outside the previously catalogued \fermi\ positional uncertainty, it is compatible with the newly determined position at the 95\% c.l.%, and is very close to the 68\% c.l. contour.
To compare our results with the position from the 4FGL-DR4 catalogue, we calculated the ratio of the areas of the error ellipses at the 95\% c.l., using the product of the two semi-axes for each ellipse. The confidence region from our analysis is smaller by a factor of $\simeq$0.7. %Additionally, we observe that the position angle of the newly determined ellipses is almost symmetrically oriented with respect to the catalogued position along the south-north direction.

Considering the large predominance of the Galactic diffuse component and the use of a weighted fit, we double-checked the goodness-of-fit using a PS map, as described by \cite{Bruel2021}.  %However, considering the densely populated nature of the region under study, the model provides an acceptable representation of the background emission.
The PS map calculation takes into account the variation of the \fermi-LAT Point Spread Function (PSF) with energy, offering a more reliable estimation of the significance with respect to the more commonly used TS map. Contrary to the TS analysis, the PS is sensitive to both positive and negative deviations from the model. Moreover, the PS map analysis is better suited for weighted likelihood fitting procedures, since it can incorporate the weights that are necessary to model the Galactic emission accurately. This PS analysis approach has been validated with previous \fermi-LAT data releases, showcasing its effectiveness. In the PS map analysis, \fermis is included in the model.
%Moreover, Figure\,\ref{fig:excess} demonstrates the absence of unmodelled contributions from background sources in the region closest to \fermis. The major excesses arise from emission along the Galactic plane and from sources with parameters fixed during our likelihood fit (i.e., these sources have been modelled according to their catalogued parameters).

Figure\,\ref{fig:PSmap} displays the PS map of the RoI, where significant detections %(PS = 2.57 corresponds to a 3-$\sigma$ detection threshold; see \citealt{Bruel2021}) 
are indicated by brighter colours. %Our source is included in the total model.
The PS map facilitates the identification of both positive and negative deviations of the model from the observed data.
We observe that the brightest sources are located close to the Galactic plane, which is marked with a dashed line. The abundance of bright areas underscores the crowded nature of the region near the Galactic plane. The inset provides a zoomed-in view of the PS map around the position of \fermis.

From the weighted PS map, we derived the histogram of the PS values (Figure\,\ref{fig:histo-PS}), which indicates the presence of some unmodelled residuals. %The histogram of PS values shown in Figure \ref{fig:histo-PS} reveals a best-fitting Gaussian distribution with a mean of 0.044$\pm$0.004 and a standard deviation of 0.607$\pm$0.008. 
To quantify the asymmetry of the residual distribution, we computed the skewness of the histogram and obtained a skewness value of 0.43. This suggests a nearly symmetrical distribution around the mean, with a slight rightward skew. The slight positive shift in the mean indicates a minor systematic tendency for the model to under-predict the data counts, though this bias remains minimal. Overall, the PS map analysis confirms that the model accurately represents the data, with few significant deviations detected.

\begin{table*}
    \centering
    \caption{Coordinates and parameters of the confidence regions for \fermis as derived from the analysis presented in this work.}
    \begin{tabular}{ccccccc}
    \hline 
    \hline
        RA & DEC & Position angle$^1$ & \multicolumn{2}{c}{68\% c.l.} & \multicolumn{2}{c}{95\% c.l.} \\
        \cmidrule(r){4-7}
        &&& Semi-major axis & Semi-minor axis & Semi-major axis & Semi-minor axis \\
    \hline 
        11$^\mathrm{h}$10$^\mathrm{m}$12.04$^\mathrm{s}$ & -65$^{\circ}$02$^{\prime}$07.4$^{\prime \prime}$& 327.2\deg & 0.054\deg & 0.045\deg & 0.087\deg & 0.073\deg \\
    \hline 
    \end{tabular}
    \begin{list}{}{}
        \item [$^1$] The position angle is measured from North to East
    \end{list}
    \label{tab:loc}
\end{table*}

\begin{figure}
\includegraphics[width=0.5\textwidth]{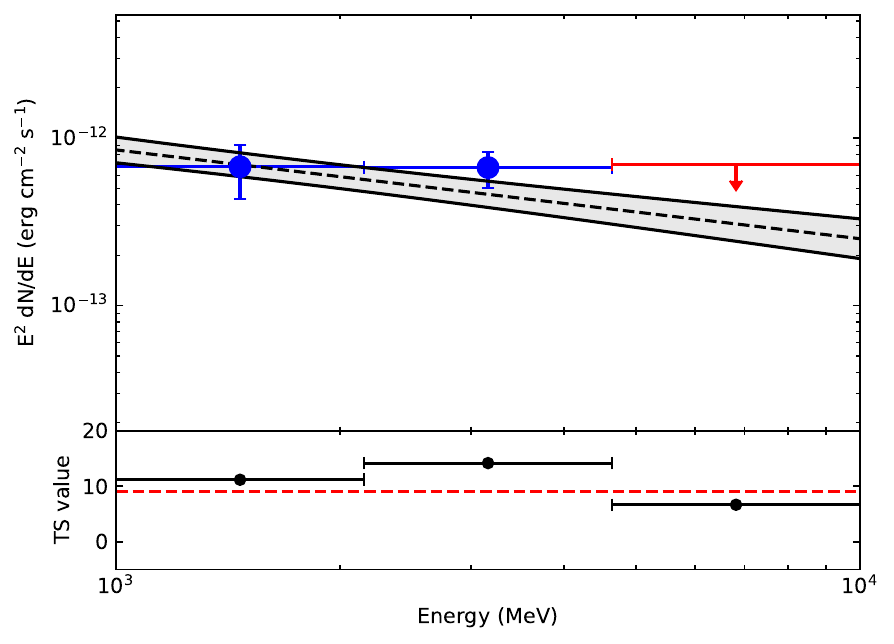}
\caption{SED with 3 equally spaced logarithmic energy bins, extracted in the 1--10\,GeV energy range. The blue data points represent detections above a 3-$\sigma$ c.l., while the red arrow represents an upper limit. The red dashed line in the bottom panel corresponds to TS = 9, meaning a 3-$\sigma$ detection level. The source is well-modelled with a power-law with photon index $\Gamma = 2.5$, here represented by the black dashed line. %. The black dashed line represents the SED model, 
The two black solid lines represent the 1-$\sigma$ bounds with respect to the model.}
\label{fig:SED}
\end{figure}

\section{Discussion}
\label{sec:discussion}
In this work, we analysed 15 years of \fermi-LAT data of the candidate tMSP \s. Our detailed study of the background and the addition of 3 more years of data led to a more accurate localisation of the gamma-ray source, demonstrating that its position is compatible with that of the candidate tMSP \s at the 95\% c.l. This supports the identification of \fermis as the gamma-ray counterpart of \s. Earlier localisations provided by the most recent \fermi\ data releases—namely the third and fourth data releases (DR3 and DR4)—which utilised 12 years of data, failed to establish a compatible localisation within the 95\% c.l. As an additional test for the gamma-ray source localisation, we performed the fitting procedure using the \chandra X-ray coordinates as the initial reference point. This yielded results consistent with our previous analysis, which employed the 4FGL-DR4 catalogued position as the starting point.

Our analysis benefits from three additional years of data compared to DR3 and DR4. We replicated the analysis procedures that led to the compilation of the DR4 catalogue by employing a joint weighted likelihood fit and incorporating all sources from the DR4 catalogue within the RoI. The extended dataset allowed for a more precise characterisation of the gamma-ray emission from \fermis and, consequently, a more accurate localisation of the source. We found that the gamma-ray source location aligns with the \chandra X-ray source \s. We anticipate that accumulating further data in the coming years will enable even more precise localisations.

%Additionally, we extracted a binned SED for \fermis. 
To compare \s to the known tMSPs, after localising the gamma-ray source, we conducted a study of its Spectral Energy Distribution (SED). Figure\,\ref{fig:SED} displays the SED extracted using three equally spaced logarithmic energy bins covering the 1--10\,GeV energy range. The dashed line represents the power-law model with the best-fitting spectral index of the source spectrum, while the shaded area indicates the 1-$\sigma$ c.l. region. The lower panel of Figure\,\ref{fig:SED} shows the TS value for each energy bin. The red dashed line marks a TS value of 9, corresponding to a $3\sigma$ detection threshold. In this analysis, we aimed to balance achieving significant detections with characterising the spectral shape. As a result, we obtained two detections above the TS = 9 threshold and an upper limit just below the threshold. %Additionally, we attempted to extract a binned light curve; however, the faintness of the source does not allow us to identify any significant variability.

The source appears relatively faint in the gamma-ray band, making it challenging to detect any time variability. The count spectrum is well-modelled by a simple power-law with a photon index of $\Gamma = 2.5 \pm 0.1$. Notably, known tMSPs (e.g., PSR J1023+0038; XSS J1227--4853; IGR J1824--2452) exhibit spectral curvature at energies below 1\,GeV \citep{Fermi4FGL, FermiDR4}. In the case of \s, identifying a similar curvature is difficult due to its faintness. Nonetheless, we tested the presence of curvature in the SED using the \texttt{curvature()} command implemented in the \textsc{Fermipy} suite. This algorithm evaluates the improvement in the fit when the source spectral model is changed from a power-law (the baseline hypothesis, corresponding to $\mathcal{L}_0$ in Eq. \ref{eq:TS}) to either a LogParabola or a power-law with a high-energy exponential cut-off (the spectral model commonly used to characterise pulsar spectra). In both cases, the TS values were below 2 ($\sim 1.1$ and $\sim 1.6$ respectively), indicating no significant curvature in the SED. Therefore, although the gamma-ray flux and photon index of our source are comparable to those of confirmed tMSPs, we cannot draw any conclusions regarding the curvature of the SED. We anticipate that additional years of data collection and further refinement of background model components may enable us to test for the presence of spectral curvature as observed in other members of this class.

\begin{figure}
    \centering
    \includegraphics[width=0.47\textwidth]{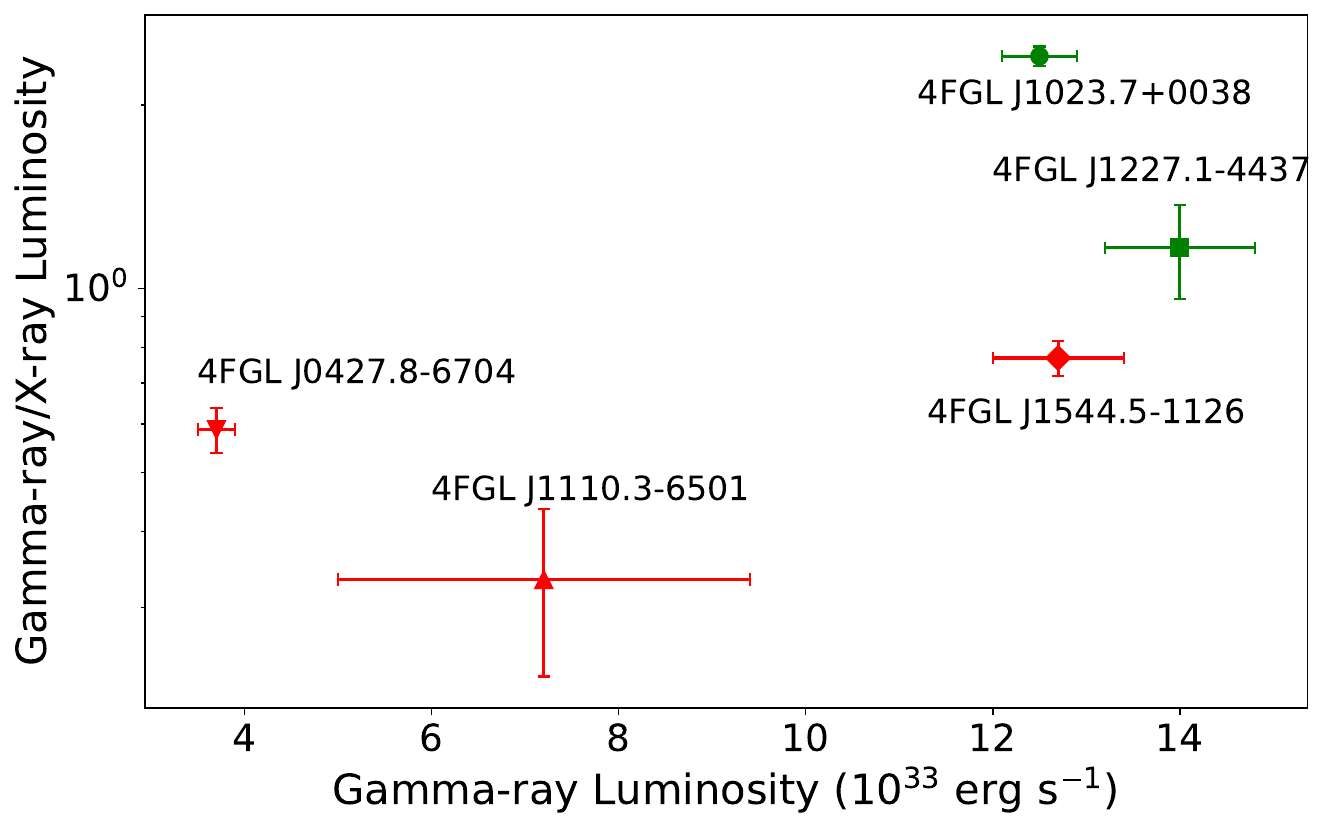}
    \caption{Gamma-ray (0.1--300 GeV) -to-X-ray luminosity (0.3--79 keV) ratio in the sub-luminous disc state as a function of the gamma-ray luminosity for known and candidate tMSPs. The green data points represent the two confirmed tMSPs with identified gamma-ray counterparts. The red data points represent candidate tMSPs, including \s. The gamma-ray luminosity for \s was calculated assuming a distance of 4\,kpc. X-ray fluxes and distances are taken from \citet{CotiZelati2019}. The gamma-ray fluxes of 4FGL J1023.7+0038 and 4FGL J1227.1--4437 are taken from \citet{Torres2017}, while the fluxes for 4FGL J1544.5--1126 and 4FGL J0427.8--6704 are estimated from the latest \fermi-LAT DR4 \citep{FermiDR4}. The gamma-ray flux of \s (\fermis in the \fermi-LAT catalogue) is derived in this work.}
    \label{fig:Lum_plot}
\end{figure}

Figure\,\ref{fig:Lum_plot} shows the ratio between gamma-ray and X-ray luminosities as a function of gamma-ray luminosity in the sub-luminous disc state. These luminosities represent average values and do not differentiate between emission modes. The luminosities for \s\ are calculated based on an assumed distance of 4\,kpc, derived from the best-fitting parallax value obtained from \emph{Gaia} data \citep{GaiaMission2016, GaiaDR32023}. We emphasise that this calculation does not account for the uncertainties associated with the parallax measurements; hence, caution should be exercised (see \citealt{CotiZelati2019} for details). Taken at face value, this estimated distance is greater than those of other known tMSPs, except for IGR J18245--2452. 
The figure reveals that the two known tMSPs with well-established gamma-ray counterparts are the brightest in the gamma-ray band among the sources examined. In contrast, \s\ exhibits the lowest gamma-ray-to-X-ray luminosity ratio, being brighter in the X-ray band than in the gamma-ray band. Establishing a correlation between these variables is challenging due to the small sample size. With a gamma-ray luminosity of $7.1 \times 10^{33}$~\unitL\ in the 0.1--300\,GeV range, \s\ is fainter than all known tMSPs, and only the candidate 4FGL J0427.8--6704 is less luminous. Notably, the latter source is also known to exhibit flaring behavior \citep{Kennedy2020}. The new estimation of the gamma-ray luminosity based on the aforementioned distance assumption is lower by a factor of 2 compared to the value estimated by \citet{CotiZelati2019}. In that study, the gamma-ray luminosity was derived from the FL8Y catalogue and was therefore considered as a preliminary estimate. Notably, the power-law index remains virtually unchanged.

The identification of a gamma-ray counterpart associated with an X-ray source that is a candidate tMSP marks a significant step in classifying the true nature of the source. However, this finding does not yet nail down \s as a tMSP, as final confirmation requires detecting a state transition between a radio pulsar state and an X-ray state (either sub-luminous or accreting outburst state). If \s\ is confirmed to be a tMSP, it would stand out as either the most distant or one of the faintest tMSPs known to date, given the currently small sample.

\section{Conclusions}
\label{sec:conclusions}
We have presented the results of the analysis of 15 years of \fermi-LAT data of the gamma-ray source \fermis. Our goal was to establish a spatial connection with the candidate tMSP \s, which is currently in an active sub-luminous X-ray state. By modeling the gamma-ray background of the region around \fermis, we detected the gamma-ray source with a TS $\sim 40$. We obtained a new location for the gamma-ray source that is more accurate than the previously catalogued position and aligns with the position derived from \chandra data at the 95\% c.l. Additionally, we extracted a binned SED that shows two significant detections above the $3\sigma$ c.l. and un upper limit. The source spectrum is modelled by a power-law with a photon index of $\Gamma = 2.5 \pm 0.1$.
%We found no signs of a cut-off energy or any curvature in the spectrum, as instead seen in other transitional millisecond pulsars in the active state \citep{Torres2022}.

Our findings underscore the critical importance of conducting detailed analyses on individual unidentified \fermi-LAT sources to accurately pinpoint their locations and conclusively identify possible counterparts at lower energies. In particular, linking an unidentified \fermi-LAT source with a variable X-ray source proves to be a highly effective method for detecting new, faint tMSPs. Following the launch of the Einstein Probe mission \citep{Yuan2022}, the research prospect in the field of tMSPs and related systems appears promising. This X-ray mission is set to conduct a comprehensive all-sky X-ray survey with unmatched sensitivity in the soft X-ray range. When combined with the expected increase in the number of unidentified gamma-ray sources detected by \fermi\ in the coming years, we anticipate the discovery of more such astrophysically compelling sources.

\begin{acknowledgements}
      We thank Nestor Mirabal for the help and suggestions with the data analysis. We thank Colin Clark, David Smith, Toby Burnett, Philippe Bruel, Xian Hou, and Matthew Kerr from the \fermi collaboration for the feedback on this work.
    We acknowledge financial contribution from the agreement ASI-INAF n.2017-14-H.0 and INAF mainstream (PI: A. De Rosa, T. Belloni), from the HERMES project financed by the Italian Space Agency (ASI) Agreement n. 2016/13 U.O and from the ASI-INAF Accordo Attuativo HERMES Technologic Pathfinder n. 2018-10-H.1-2020. We also acknowledge support from the European Union Horizon 2020 Research and Innovation Frame- work Programme under grant agreement HERMES-Scientific Pathfinder n. 821896 and from PRIN-INAF 2019 with the project "Probing the geometry of accretion: from theory to observations" (PI: Belloni).

    FCZ, AMarino and NR are supported by the H2020 ERC Consolidator Grant “MAGNESIA” under grant agreement No. 817661 (PI: Rea) and from grant SGR2021-01269 (PI: Graber/Rea).
    FCZ is supported by a Ram\'on y Cajal fellowship (grant agreement RYC2021-030888-I). This work was also partially supported by the program Unidad de Excelencia Maria de Maeztu CEX2020-001058-M. DFT acknowledges support from grants PID2021-124581OB-I00 funded by MCIN/AEI/10.13039/501100011033 and SGR2021-00426.

    The Fermi-LAT Collaboration acknowledges support for LAT development, operation and data analysis from NASA and DOE (United States), CEA/Irfu and IN2P3/CNRS (France), ASI and INFN (Italy), MEXT, KEK, and JAXA (Japan), and the K.A. Wallenberg Foundation, the Swedish Research Council and the National Space Board (Sweden). Science analysis support in the operations phase from INAF (Italy) and CNES (France) is also gratefully acknowledged. This work performed in part under DOE Contract DE-AC02-76SF00515.
\end{acknowledgements}

\bibliographystyle{aa}
\bibliography{bibliography}

\end{document}